\documentclass[12pt,preprint]{aastex}

\newcommand{\etal}{et al.}

\shorttitle{AGN and the CRB}
\shortauthors{Draper, Northcott, \&  Ballantyne}

\begin{document}

\title{Are Active Galactic Nuclei the Solution to the Excess Cosmic Radio Background at 1.4 GHz?}


\author{A. R. Draper, S. Northcott, and D. R. Ballantyne}
\affil{Center for Relativistic Astrophysics, School of Physics,
  Georgia Institute of Technology, Atlanta, GA 30332}
\email{aden.draper@physics.gatech.edu}

\begin{abstract}
Recently the ARCADE 2 experiment measured the cosmic radio background (CRB) and 
found the brightness temperature of the CRB at 1.4 GHz to be $\sim$480 mK.   
Integrating the flux density from the observed 1.4 GHz radio source count produces a brightness temperature of 
$\sim$100 mK---less than a quarter of the observed CRB at 1.4 GHz.  Radio quiet AGN are 
a large fraction of the 1.4 GHz $\mu$Jy sources and typically host significant star formation.  Thus, it 
is possible that AGN and host star formation could be responsible for some fraction of the excess CRB at 
1.4 GHz.  Here, an X-ray background population synthesis model is used in conjunction with empirical radio 
to X-ray luminosity ratios to calculate the AGN contribution to the CRB at 1.4 GHz including the emission 
from host star formation.  It is found that AGN and host star formation contribute $\lesssim$9$\%$ of the CRB at 1.4 GHz.  
When all known 1.4 GHz radio source classes are considered, $\lesssim$60$\%$ of the CRB at 1.4 GHz 
is accounted for; therefore, it is necessary that either known radio sources evolve significantly at flux densities 
below current survey sensitivity limits or a new population of low flux density radio sources exist.


\end{abstract}



\section{Introduction}
\label{sect:intro}

Extragalactic backgrounds are important tools in understanding the density and evolution of different objects within the universe.  
The $\gamma$-ray background places constraints on dark matter annihilation \citep[e.g.,][]{C10}, the X-ray 
background (XRB) provides a census of accretion onto supermassive black holes \citep[e.g.,][]{G07}, the cosmic infrared 
background (CIRB) encodes the history of obscured star formation and galaxy evolution \citep[e.g.,][]{Lutz11}, and 
the cosmic microwave background (CMB) provides the deepest insights into the early universe \citep[e.g.,][]{K11}.  
It is known that at least two source populations contribute to the cosmic radio background (CRB).  At fluxes $>$ 1 mJy
radio loud active galactic nuclei (AGN), actively accreting supermassive black holes defined by $f_{5}$/$f_B$ $\gtrsim$ 
10, where $f_5$ is the 5 GHz flux density and $f_B$ is the B band flux density \citep{K89}, are the dominate source population \citep[e.g.,][]{W93}.  
At lower flux densities, the 1.4 GHz source count steepens as star forming galaxies (SFGs) and radio quiet AGN begin to 
dominate.  Recently, the ARCADE 2 experiment measured the CRB at frequencies between 3 and 90 GHz and found that the spectrum of the
observed CRB suggests that the extragalactic brightness temperature at 1.4 GHz
is $\sim$480 mK (Fixsen et al. 2011, their eq. 6).  Even when considering the large dispersion between various 
observed source counts, extrapolating the total observed source count at 1.4 GHz to lower 
flux densities can only account for $\sim$100 mK of the observed CRB \citep{G08,S11,V11}.  The next step 
must be to consider phenomenological models of individual source classes.  
Here, the contribution of AGN and their hosts to the CRB is investigated. 


Radio quiet AGN, although less radio luminous, are much more common than radio loud AGN, and thus are an important 
class of sources in the $\mu$Jy flux density range at 1.4 GHz \citep[e.g.,][]{P09}.  Furthermore, radio 
quiet AGN host galaxies are known to host significant star formation \citep[e.g.,][]{L10,Y10}.  Therefore, it is 
plausible that emission from radio quiet AGN and their hosts makes a significant contribution to the CRB at 1.4 GHz.  Using 
optical AGN luminosity functions, \citet{S10} find that radio quiet AGN should contribute a 
few percent of the CRB.  While optical AGN luminosity functions do include emission from the host galaxy \citep{DB11b}, star 
formation in AGN hosts tends to be obscured \citep[e.g.,][]{L10}; therefore, AGN host star formation contributes very little 
to the total optical luminosity of the AGN system.  Star formation in AGN hosts must be carefully considered in order to 
understand the possible range of CRB fractions which can be accounted for by AGN system radio emission.

Here, a detailed accounting of the AGN contribution to the CRB is conducted with the inclusion of star formation in  
AGN host galaxies.  The framework of XRB population synthesis modeling is combined with empirical X-ray to radio flux 
ratios.  This allows for the consideration of AGN with radio flux densities well below 
the current survey sensitivity limits.  Variations in the AGN contribution to the CRB due to uncertainty in the evolution of AGN host star formation, the 
AGN luminosity function, and the AGN radio to X-ray luminosity ratio are also explored.  A 
$\Lambda$-dominated cosmology is assumed with $h$ = 0.7, and $\Omega_{\Lambda}$ = $1-\Omega_{m}$ = 0.7 \citep{Sperg03}.  
Radio sources have a flux density with spectral index $\alpha$, such that $S$ $\propto$ $\nu^{\alpha}$.

\section{Calculations}
\label{sect:calc}

The AGN number counts at 1.4 GHz are calculated between $S_{max}$ = 10 mJy and $S_{min}$ = 100 nJy, in the same manner as by \citet{B09}.  
An XRB synthesis model is used to characterize the space density of AGN, 
the AGN type 2/type 1 ratio, $f_2$, and the fraction of Compton thick (CT) AGN, $f_{CT}$ 
\citep[e.g.,][]{B06,DB09}, AGN with obscuring column densities $N_H$ $>$ 10$^{24}$ cm$^{-2}$.  Either the 
\citet{U03} or \citet{LF05} hard X-ray luminosity function (HXLF) is used to model the space density and 
evolution of AGN.  The evolution of $f_2$ is dependent on both $L_X$, the AGN 2--10 keV luminosity, 
and redshift, $z$, in the manner determined by \citet{B06}.  The $f_{CT}$ is determined by fitting the 
peak of the XRB at $\sim$ 30 keV and the local space density of CT AGN \citep{DB10, B11}. Both the scenario 
where CT AGN evolve like less highly obscured type 2 AGN (10$^{22}$ $<$ $N_H$ $<$ 10$^{24}$ cm$^{-2}$) and 
the scenario where CT AGN are in specific evolutionary states are considered \citep[see][]{DB10}.

Following \citet{TW03}, the ratio of 5 GHz radio luminosity to $L_X$ for radio quiet AGN, 
defined as $R_X$ $\equiv$ $\log (\nu L_{\nu}(5 GHz)/L_X)$ is characterized as
\begin{equation}
\label{eq:tw}
R_X = \left \{ \begin{array}{ll}
            -0.67\log L_X + 23.67 & 41.5 \leq \log L_X \leq 43\\
             -5 & 43 < \log L_X \leq 44 \\
            \log L_X -49 & 44 < \log L_X \leq 45 \\
             -4 & \log L_X > 45.
            \end {array}
      \right.
\end{equation}
The AGN 5 GHz luminosity is converted to 1.4 GHz using a spectral index of $\alpha$ = -0.7\footnote{If the $\alpha$ distribution for radio quiet AGN observed by \citet{K98} is approximated as Gaussian with $\bar{\alpha}$ = -0.7 and $\sigma$ = 0.6, the radio quiet AGN contribution to the brightness temperature increases by $<$ 1 mK.} \citep{K98}.  \citet{B09} showed that the radio quiet AGN $R_X$ relation 
of \citet{P07} is also consistent with the current observed AGN radio number counts.  
The \citet{P07} radio quiet $R_X$, based on observations of local Seyferts, is nearly constant with luminosity.
Radio 
loud AGN X-ray luminosities are converted to radio luminosities using the relation 
\begin{equation}
\label{eq:rl}
R_X \approx -2.0
\end{equation}
\citep{TW03}.  For the radio loud AGN a spectral index of 
$\alpha$ = -0.8\footnote{If the $\alpha$ distribution for radio loud AGN observed by \citet{LM07} is approximated as Gaussian with $\bar{\alpha}$ = -0.8 and $\sigma$ = 0.6, the radio loud AGN contribution to the brightness temperature decreases by $<$ 0.5 mK.} is used to convert from 5 GHz to 1.4 GHz luminosity.  

The HXLF contains both radio quiet and radio loud AGN.  In 
order to separate the radio quiet AGN and the radio loud AGN, a radio loud fraction, $f_{RL}$, must be 
defined.  \citet{B09} found 
that radio number counts between 1 and 10 mJy were well fit assuming an exponential increase from $f_{RL}$ = 0.0175 at 
$\log L_X$ = 41.5 to $f_{RL}$ = 0.10 at $\log L_X$ $\geq$ 46.0, at all $z$.  This is consistent 
with the overall radio loud fraction being 0.1-0.2 \citep[e.g.,][]{D09}.

It is well documented that galaxies hosting AGN also tend to host on-going star formation 
\citep[][and references therein]{DB11}, thus the radio flux density due to star formation within AGN host galaxies must be considered 
for an accurate calculation of the AGN radio number counts.  The \citet{B03} calibration is 
used to convert a star formation rate (SFR) into a 1.4 GHz radio luminosity assuming that the infrared-radio correlation (IRC) 
does not evolve with $z$.  For the radio spectrum 
of star forming regions, $\alpha$ = -0.8 is assumed \citep{Y01}.  The 1.4 GHz radio differential 
number counts, $dN/dS$, are then calculated as in Equation 2 of \citet{B09}.

Once $dN/dS$ is known, calculation of the brightness temperature, $T(\nu)$, is straight forward.  
The intensity $I(\nu)$ is related to $dN/dS$ such that
\begin{equation}
\label{eq:int}
I(\nu) = \int^{S_{max}}_{S_{min}} \frac{dN}{dS}(\nu)\cdot S dS.
\end{equation}
The Rayleigh-Jeans approximation gives 
\begin{equation}
\label{eq:temp}
T(\nu) = I(\nu)\frac{\lambda^2}{2k},
\end{equation}
where $\lambda$ = 21 cm is the wavelength and $k$ is the Boltzmann constant.  
Thus, by applying the machinery of the XRB and using empirically determined 
conversions between X-ray and radio luminosities, the AGN contribution to the 
1.4 GHz CRB is calculated.

With the basic model in place, the contribution of AGN systems to the CRB is investigated.  
The results from the \citet{U03} and \citet{LF05} HXLFs and the \citet{TW03} and 
\citet{P07} radio quiet $R_X$ relations are compared.  Constant and evolving star 
formation laws are explored.  Both an evolving and non-evolving model of $f_{CT}$ 
are considered.  Analyzing such a large portion of the parameter space allows the 
maximum possible contribution to the CRB from AGN to be determined. 
The resulting models are then constrained by the observed 1.4 GHz number counts.

\section{Results}
\label{sect:res}

\subsection{Bare AGN}
\label{sub:bare}

\begin{figure*}
\begin{center}
\includegraphics[angle=0,width=0.95\textwidth]{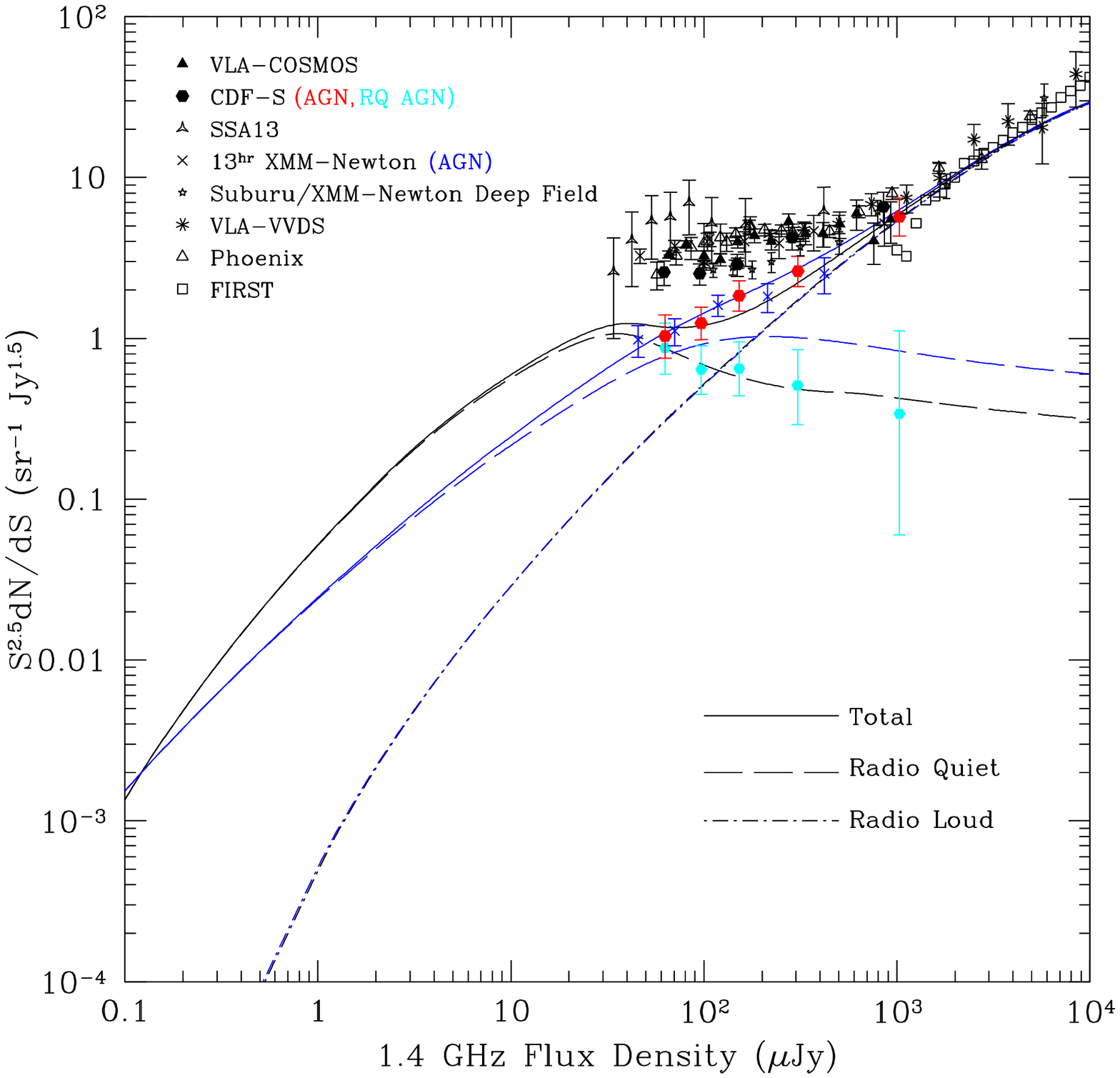}
\end{center}
\caption{Euclidean normalized 1.4 GHz differential number counts for bare AGN and AGN with star formation.  The blue lines show bare AGN assuming the \citet{LF05} HXLF and \citet{P07} $R_X$, as summarized in the fourth line of Table \ref{crbagn}.  The black lines show AGN with host SFR as described by Equation \ref{eq:sfr}, the \citet{LF05} HXLF and the \citet{TW03} $R_X$, as summarized in the tenth row of Table \ref{crbagn}.  Also shown are the observed source counts from a variety of surveys: VLA-COSMOS \citep{B08}, CDF-S \citep{K08}, SSA 13 \citep{F06}, 13 hr {\em XMM-Newton} \citep{S04}, Suburu/{\em XMM-Newton} Deep Field \citep{S06}, VLA-VVDS \citep{Bo03}, Phoenix \citep{H03}, and FIRST \citep{W97}.  The blue and red points show the estimated AGN radio counts from the 13 hr {\em XMM-Newton} \citep[blue;][]{S08} and CDF-S \citep[red;][]{Padplus11}.  The cyan points show the estimated radio quiet AGN 1.4 GHz counts in the CDF-S \citep{Padplus11}.}
\label{fig:agn}
\end{figure*}

First, the host star formation is ignored and the contribution to the CRB from bare AGNs is 
considered and shown in the upper part of Table \ref{crbagn}.  
The listed $\chi_{red}^2$ values refer to the 15 AGN data points from \citet{S08} and 
\citet{Padplus11} shown as the red, blue, and cyan data in Figure \ref{fig:agn}.    
The \citet{P07} $R_X$ predicts an AGN brightness 
temperature $\sim$5 mK higher than the \citet{TW03} $R_X$.  Similarly, the \citet{LF05} 
HXLF produces an AGN brightness temperature $\sim$5 mK higher than 
the brightness temperature predicted by the \citet{U03} HXLF.  The blue 
lines in Figure \ref{fig:agn} show the Euclidean normalized 1.4 GHz differential number counts 
for the bare AGN model with the lowest $\chi_{red}^2$, which assumes the \citet{LF05} HXLF 
and the \citet{P07} $R_X$. Thus, bare AGN appear to only contribute 
$\sim$4$\%$ of the CRB at 1.4 GHz, in agreement with the results of \citet{S10}.


The minimum X-ray luminosity of the XRB framework is $L_X^{min}$ = 10$^{41.5}$ erg s$^{-1}$, which 
corresponds to a minimum radio loud AGN 1.4 GHz radio luminosity of 1.8 $\times$ 10$^{23}$ W Hz$^{-1}$.  In the 
VLA-COSMOS survey, radio loud AGN were observed with 1.4 GHz radio luminosities down to 
$\sim$5 $\times$ 10$^{21}$ W Hz$^{-1}$ \citep{S09}.  Therefore, the 1.4 GHz radio loud AGN luminosity 
function of \citet{S09} is used to assess the importance of the low luminosity radio loud AGN.  The 
radio loud AGN contribution to the CRB using both the pure density evolution 
(PDE) and pure luminosity evolution (PLE) versions of the \citet{S09} luminosity function are shown in the 
fifth and sixth rows of Table \ref{crbagn}. The dot-dashed line in Figure \ref{fig:max} shows the radio loud 
contribution to the differential number counts using the \citet{S09} PDE luminosity function.  As radio loud 
AGN are X-ray luminous for only a fraction of their lifetime \citep{DB09}, 
it is possible to assume that all X-ray selected AGN in the \citet{U03} HXLF are radio quiet AGN.
In this case, bare AGN contribute $\sim$6--8$\%$ of the CRB at 1.4 GHz, depending on whether the PLE 
or PDE evolution is used for the radio loud luminosity function.

\subsection{Accounting for Star Formation}
\label{sub:sf}

It is known that AGN hosts have star formation \citep{DB11}; thus, we now consider the 
effect of AGN host star formation on the AGN and host contribution to the CRB.  
\citet{B09} found that if AGN have either SFR 
$\approx$ 2--3 M$_{\odot}$ yr$^{-1}$ or 
\begin{equation}
\label{eq:sfr}
SFR \approx 0.25(1+z)^{1.76}(\log L_X -40)^{3.5} M_{\odot}\ yr^{-1},
\end{equation}
the predicted number counts are in good agreement with observations, as shown by the black lines in Figure \ref{fig:agn}.  
However, as shown in the bottom half of Table \ref{crbagn}, adding star 
formation to AGN hosts only increases the contribution of AGN and their hosts to the CRB by $\lesssim$10 mK. 

We also consider the evolving CT AGN model of \citet{DB10}, in which CT AGN have a physically motivated 
Eddington ratio distribution.  This model is in agreement with the XRB, the local space density of CT 
AGN, and $z$ $\sim$ 2 mid-infrared estimations of the CT AGN space density \citep{DB10}.  
The third to last row in Table \ref{crbagn} shows 
the AGN contribution to the CRB using the evolving $f_{CT}$ model and the evolving SFR of \citet{B09}.  The evolving 
model is in better agreement with the observed number counts and results in a higher brightness temperature than the non-evolving model.

\begin{figure*}
\begin{center}
\includegraphics[angle=0,width=0.95\textwidth]{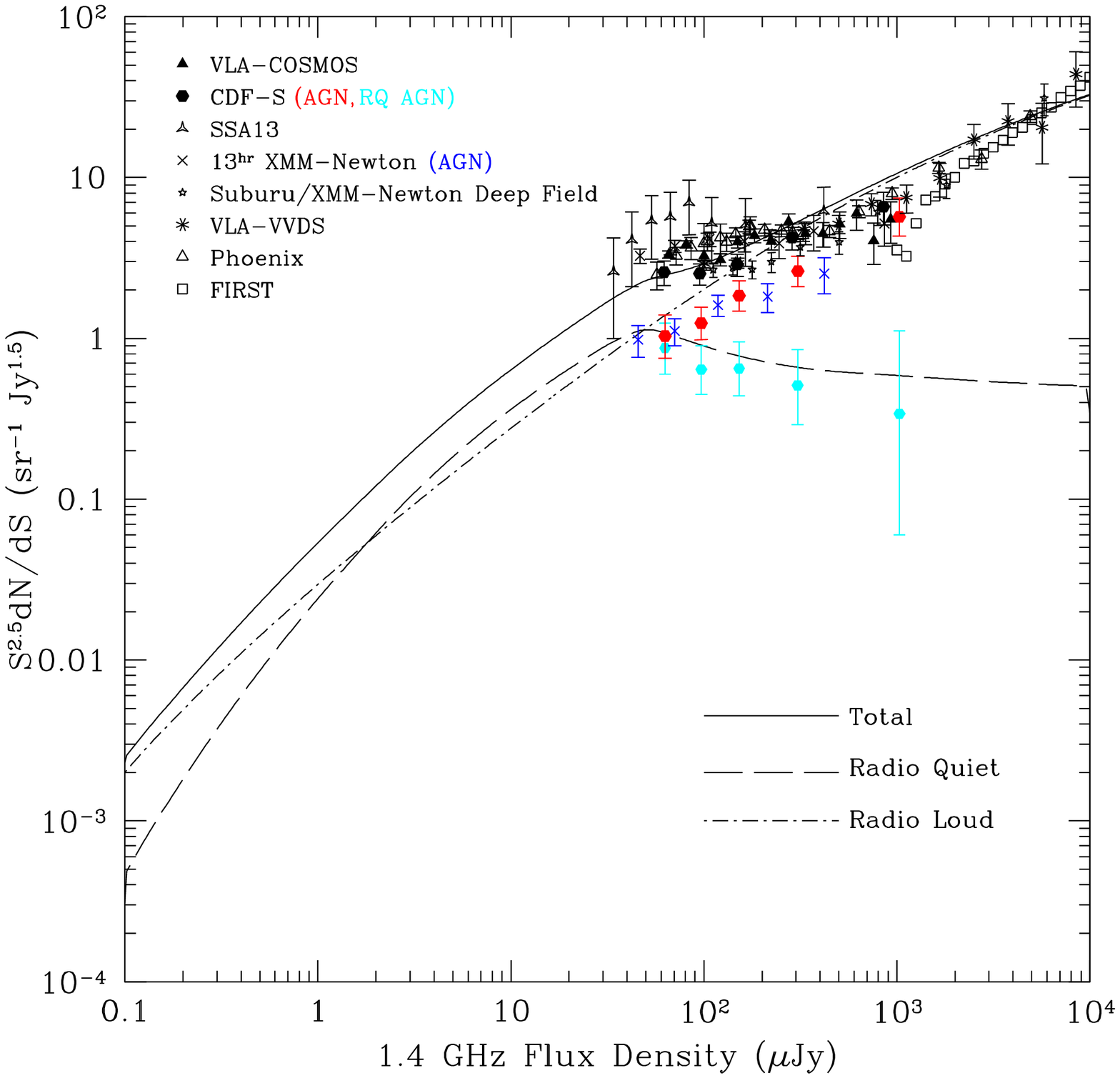}
\end{center}
\caption{Euclidean normalized differential number counts for the maximal AGN and host star formation contribution to the 1.4 GHz CRB (last line of Table \ref{crbagn}).  The radio loud AGN are accounted for using the \citet{S09} PDE luminosity function and the radio quiet AGN are accounted for using the \citet{U03} HXLF, the evolving $f_{CT}$ model of \citet{DB10}, and the \citet{TW03} $R_X$.  The SFR of the radio quiet AGN hosts is described by Equation \ref{eq:sfr}.  This model gives the maximal total AGN and host brightness temperature of 0.042 K.  The plotted data is the same as in Figure \ref{fig:agn}.}
\label{fig:max}
\end{figure*}

In order to evaluate the maximum possible contribution of AGN and their hosts to the CRB, it is 
assumed that AGN systems dominate the 1.4 GHz radio number counts down to the 10 $\mu$Jy flux density level.  
For this scenario, the radio loud AGN are accounted for using the 
\citet{S09} PDE luminosity function, and it is assumed that all AGN in the \citet{U03} HXLF are 
radio quiet.  The evolving $f_{CT}$ model is used and the radio quiet AGN hosts are assigned the 
SFR described in Equation \ref{eq:sfr}.  This maximal scenario is shown in Figure \ref{fig:max} 
and summarized in the last row of Table \ref{crbagn}.
Thus, it is clear that AGN contribute to the CRB; however, the maximum 
contribution to the CRB from AGN and their hosts, 
which is in reasonable agreement with the observed source count, as shown in Figure \ref{fig:max}, is 
$\sim$42 mK, or $\sim$9$\%$, of the CRB at 1.4 GHz.  However, in the best fit model, as assessed 
by the minimum $\chi_{red}^2$, AGN and their hosts contribute only $\sim$28 mK, or $\sim$6$\%$, of the 
CRB at 1.4 GHz.

\section{Discussion and Summary}
\label{sect:disc}

We have shown that AGN systems contribute $\lesssim$9$\%$ of the CRB at 1.4 GHz.  
Thus the contribution of other source classes to the CRB must be considered.  In Table 
\ref{crb} the known dominant source classes of the CRB are listed with the expected 
contribution of each source class. 

It is known that star forming galaxies (SFGs) are an important source class in the $\mu$Jy and nJy flux density ranges.  
Integrating the number counts predicted by the \citet{Padplus11} 1.4 GHz SFG luminosity function suggests that SFG only contribute 
$\sim$3$\%$ of the CRB at 1.4 GHz.  However, current surveys do not reach radio flux densities $<$ 10 $\mu$Jy; thus the 
best estimates of the contribution of SFGs to the CRB, which do not require extrapolation of the observed source counts, 
are computed using the IRC and the CIRB.  If the IRC does not evolve with $z$ and maintains the locally observed 
correlation at high $z$, then SFGs only contribute $\sim$9$\%$ of the CRB at 1.4 GHz \citep{P11}.  It is unclear 
whether the IRC evolves with $z$ or not, however, \citet{P11} find that if the IRC evolves with $z$ in a manner which is 
consistent with current observations, SFGs still only contribute $\sim$14$\%$ of the CRB at 1.4 GHz.  It is possible 
that low power, high redshift AGN systems, which observationally would be difficult to distinguish from SFGs, may increase 
this percentage \citep{S10}. 
  
As AGN, based on the calculation presented here, and SFGs, based on estimations from the IRC, can account for only $\sim$25$\%$ 
of the CRB at 1.4 GHz, additional low flux density source classes 
must be considered.  One possible source class is radio supernovae (RSNe), however, these objects are short-lived 
and not common enough to make a significant contribution to the CRB \citep{S10}.  \citet{Pad11} suggests that low 
radio power ellipticals (LRPEs) and dwarf galaxies may be important source classes at flux densities currently below survey 
sensitivities.  LRPEs are predicted to have similar surface densities as AGN, and thus are not a 
dominate contributor to the CRB.  Dwarf galaxies are expected to dominate the source count at nJy flux densities, 
but \citet{Pad11} predicts that dwarf galaxies will only contribute $\lesssim$8$\%$ of the CRB.  Furthermore, \citet{S11} 
point out that for a population of sub-mJy point sources to account for the difference between the observed CRB and the 
brightness temperature predicted from the observed source count, the surface 
density of this population must exceed the surface density of galaxies in the Hubble Ultra Deep Field \citep{Beck06}.  
Therefore, it is likely that the unaccounted for CRB is due to the evolution of known sources or an evolution of processes, 
such as those responsible for the IRC, and not due to an unknown source class.

Despite the current lack of a dominant candidate point source class, it is unlikely the 1.4 GHz CRB is dominated by 
diffuse sources.  \citet{S10} considers the contribution from low-surface-brightness sources.  Sources which may 
emit a large flux over an extended area may have low enough surface brightness to be missed by current survey source 
counts.  \citet{S10} finds that the low-surface-brightness sources can at most contribute $\sim$25$\%$ of the CRB.  
Furthermore, the constraints placed on the CRB from the XRB and $\gamma$-ray background are inconsistent with the 
CRB being dominated by diffuse emission from a population of relativistic electrons within the intracluster medium 
or intergalactic medium \citep{S10, L11}.  Thus, other sub-$\mu$Jy sources must be considered in order to fully account 
for the CRB at 1.4 GHz.

In conclusion, a detailed accounting of the AGN and host star formation contribution to the CRB has been carried out using the constraints of 
XRB population synthesis modeling and empirical X-ray to radio AGN luminosity ratios.  This study has shown that AGN and host star formation
contribute $\lesssim$9$\%$ of the CRB at 1.4 GHz.  In a best case 
scenario, all sources currently considered in the literature contribute $\lesssim$60$\%$ of the CRB at 1.4 GHz.  The 
unaccounted for radio emission must come from either a new low flux density 
radio population or from a known radio population with a 
strong evolution of the radio emission for low flux density objects.  Therefore, it is 
necessary that radio surveys continue to explore the radio source count down to sub-$\mu$Jy levels in order to resolve the 
source class responsible for the unaccounted for CRB flux density at 1.4 GHz.

\acknowledgments
This work was supported by NSF award AST 1008067.  The authors thank the anonymous referee for helpful comments.


{}



%
\begin{deluxetable}{lcccc}
\tablecolumns{5}
\tablecaption{Contribution of AGN to the 1.4 GHz CRB}
\tablehead{
  \colhead{HXLF} &
  \colhead{Star Formation Law} &
  \colhead{T$_{total}$ (K)} &
  \colhead{T$_{RL}$ (K)} &
  \colhead{$\chi^2_{red}$} \\
}
\startdata
    \cutinhead{Bare AGN}
    U03 & - & 0.013 & 0.011 & 8.3 \\
    U03\tablenotemark{a} & - & 0.019 & 0.011 & 1.0 \\
    LF05 & - & 0.018 & 0.015 & 4.1 \\
    LF05\tablenotemark{a} & - & 0.023 & 0.015 & 0.6 \\
    S09 PDE + U03\tablenotemark{b} & - & 0.033 & 0.031 & 7.6 \\
    S09 PLE + U03\tablenotemark{b} & - & 0.023 & 0.021 & 1.7 \\

    \cutinhead{AGN + Star Formation}
    U03 & SFR = 3 M$_{\odot}$ yr$^{-1}$ & 0.021 & 0.011 & 1.5 \\
    LF05\tablenotemark{a} & SFR = 2 M$_{\odot}$ yr$^{-1}$ & 0.028 & 0.015 & 1.3 \\
    U03 & Eq. \ref{eq:sfr} & 0.019 & 0.011 & 1.9 \\
    LF05 & Eq. \ref{eq:sfr} & 0.028 & 0.015 & 0.4 \\
    U03\tablenotemark{c} & Eq. \ref{eq:sfr} & 0.024 & 0.014 & 0.9 \\
    U03\tablenotemark{c} & Eq. \ref{eq:sfr}\tablenotemark{d} & 0.029 & 0.014 & 5.5 \\
    S09 PDE + U03\tablenotemark{b,c} & Eq. \ref{eq:sfr} & 0.042 & 0.031 & 19 \\

\enddata
\tablenotetext{a}{Radio quiet AGN $L_X$ is converted to radio luminosity using the \citet{P07} conversion.  For all other models the radio quiet $L_X$ is converted to radio luminosity using Equation \ref{eq:tw}.}
\tablenotetext{b}{Radio loud AGN are accounted for using the \citet{S09} 1.4 GHz radio luminosity function for radio loud AGN which is evolved using either pure density evolution (PDE) or pure luminosity evolution (PLE).  The radio quiet AGN are modeled using the \citet{U03} HXLF assuming that all X-ray AGN are radio quiet.}
\tablenotetext{c}{Uses the evolving $f_{CT}$ model of \citet{DB10}}
\tablenotetext{d}{Star formation follows the functional form of equation \ref{eq:sfr} with a different normalization factor.  Compton thin AGN have normalization factor 0.3 while low Eddington ratio CT AGN have normalization factor 0.05 and high Eddington ratio CT AGN have normalization factor 4.0}
\tablecomments{U03 refers to \citet{U03}, LF05 refers to \citet{LF05}, and S09 refers to \citet{S09}.  T$_{RL}$ is the brightness temperature of radio loud AGN.}
\label{crbagn}
\end{deluxetable}
\begin{deluxetable}{clcl}
\tablecolumns{4}
\tablecaption{Contributions of various sources to the 1.4 GHz CRB}
\tablehead{
  \colhead{ } &
  \colhead{ } &
  \colhead{Brightness Temperature (K)} &
  \colhead{Reference} \\
}
\startdata
     & Total Measured CRB & 0.48 $\pm$ 0.07 & \citet{F11} \\
    \hline \\
     & AGN & 0.018 & this work \\
     & AGN+SF & 0.025 & this work \\
    (1) & Max AGN+SF & 0.042 & this work \\
    & & & \\
     & SFG & 0.015\tablenotemark{a} & this work \\
     & SFG non-evolving IRC & 0.040\tablenotemark{b} & \citet{P11} \\
    (2) & SFG evolving IRC & 0.063\tablenotemark{c} & \citet{P11} \\
     & & \\
    (3) & RSNe & $\lesssim$ 0.00017 & \citet{S10} \\
    (4) & LRPEs & 0.010\tablenotemark{d} & \citet{Pad11} \\
    (5) & Dwarf SFGs & 0.038 & \citet{Pad11} \\
    (6) & Low-surface-brightness Sources & 0.120 & \citet{S10} \\
     & & & \\
     & Total (1--6) & 0.27 & \\
\enddata
\tablenotetext{a}{Calculated by integrating the number counts predicted by the \citet{Padplus11} observed SFG broken power law luminosity function at 1.4 GHz and  assuming $\alpha$ = -0.8.}
\tablenotetext{b}{Converted from 1 GHz using Eq. 19 and 20 of \citet{P11}.}
\tablenotetext{c}{Converted from 1 GHz using Eq. 19 and 21 of \citet{P11}.}
\tablenotetext{d}{Estimated based on the expectation that LRPEs have surface density and flux densities similar to radio quiet AGN.}

\tablecomments{SF refers to star formation, SFG refers to star forming galaxy, IRC refers to the infrared-radio correlation, RSNe refers to radio supernovae, and LRPE refers to low radio power ellipticals.}

\label{crb}
\end{deluxetable}

\end{document}